# Nanoscale Confinement Enhances Ultrafast Demagnetization

*Yoav William Windsor, Tobias Lojewski, Moumita Kundu, Klaus Sokolowski-Tinten, Nico Rothenbach, Andrea Eschenlohr, Markus Ernst Gruner, Katharina Ollefs, Carolin Schmitz-Antoniak, Soma Salamon, Daniela Zahn, Laurenz Rettig, Christian Schüßler-Langeheine, Niko Pontius, Renkai Li, Mianzhen Mo, Suji Park, Xiaoshe Shen, Stephen Weathersby, Xijie Wang, Rossitza Pentcheva, Heiko Wende, Ulrich Nowak, and Uwe Bovensiepen*

Y. W. Windsor
Institut für Physik und Astronomie, Technische Universität Berlin, Straße des 17. Juni 135, 10623 Berlin, Germany

Y. W. Windsor, D. Zahn, L. Rettig
Fritz Haber Institute of the Max Planck Society, Faradayweg 4-6, 14195 Berlin, Germany

T. Lojewski, K. Sokolowski-Tinten, N. Rothenbach, A. Eschenlohr, M. E. Gruner, K. Ollefs, S. Salamon, X. Wang, R. Pentcheva, H. Wende, U. Bovensiepen
Faculty of Physics and Center for Nanointegration Duisburg-Essen (CENIDE), University of Duisburg-Essen, Lotharstr. 1, 47057 Duisburg, Germany

M. Kundu, U. Nowak
Department of Physics, University of Konstanz, 78457 Konstanz, Germany

C. Schmitz-Antoniak
Technical University of Applied Sciences Wildau, Hochschulring 1, 15745 Wildau, Germany

C. Schüßler-Langeheine, N. Pontius
Helmholtz-Zentrum Berlin für Materialien und Energie GmbH, Albert-Einstein-Str. 15, 12489 Berlin, Germany

R. Li, M. Mo, S. Park, X. Shen, S. Weathersby, X. Wang
SLAC National Accelerator Laboratory, 2575 Sand Hill Rd., Menlo Park, California 94025, USA

X. Wang
Department of Physics, TU Dortmund University, 44227 Dortmund, Germany

U. Bovensiepen
Institute for Solid State Physics, The University of Tokyo, Kashiwa, Chiba 277-8581, Japan

Current address of D. Zahn:
Fraunhofer Institute EMFT, Hansastraße 27D, 80686 Munich, Germany

**Nanoscale miniaturization has revolutionized the field of spintronics by enabling exponential growth in areal bit density. A similar leap is also expected in device speeds through successfully harnessing femtosecond magnetization dynamics. However, combining this with the miniaturization of realistic devices is challenging. To address this, we studied the effect of dimensional confinement on the femtosecond demagnetization of Fe. By gradually increasing the level of confinement while keeping excitation conditions constant, we found that Fe layers thinner than 10 nm exhibit enlarged demagnetization amplitudes, reaching a ~75% increase at 2 nm. By combining ultrafast experiments sensitive to the spins, the charge carriers, and the phonons, we establish that this finite-size effect is magnetic in origin and is not phonon-driven. With the support of *ab-initio* calculations and atomistic spin dynamics simulations, we identify the enhancement effect as due to local weakening of spin order at the Fe's interface, which becomes significant upon increased confinement.**

## 1. Introduction

The ever-growing demand for information technology fuels the advancement of technological frontiers, such as miniaturization, device speed, and energy efficiency. The "Moore's law" style of exponential growth in areal bit density over the last decades may be extended through advances in the field of spintronics, which has seen tremendous progress in miniaturization. [1] This has been largely due to advances in synthesizing nanometer-scale ferromagnetic (FM) layers.[2] For advances in speed, [3] the field has seen intense research into femtosecond spin dynamics since ultrafast demagnetization was first reported nearly three decades ago.[4–6] Reaching such device speeds would enable computation well beyond THz clock speeds. [7]

However, practical utilization of ultrafast demagnetization has been limited to date, particularly in electronic devices. An important question that remains unanswered is whether this effect can be realistically harnessed without compromising the advances already made in miniaturization. In other words, how is femtosecond demagnetization affected by the extreme reductions in size that enable high bit density? And are there fundamental limits on the dimensions of a ferromagnet that needs to be both stable enough for technological applications, but also susceptible enough to react to femtosecond laser pulses? Reliably addressing such questions is tricky, [8] because varying the amount of material on such small scales can significantly affect measurement accuracy, such as due to variations in optical penetration depths which can affect pump excitation densities.

Here we address these questions in elemental Fe, using a specific sample design strategy that overcomes such issues. We study a series of thin samples that all contain an identical amount of Fe, but with a different number of electrically insulating MgO spacer layers within them (Figure 1a). The spacers are transparent to our optical excitations, and our Fe $L_3$-edge X-ray probe is only resonant to the Fe. In this way, we effectively probe ultrafast demagnetization of Fe under varying levels of confinement without the need to consider variations in the excitation. Already in equilibrium conditions nm-scale dimensional confinement is known to affect FM magnetization. [9–11] We find that the amplitude of ultrafast demagnetization is significantly enhanced when the Fe is confined to thicknesses below $d_{Fe} \approx 10\ nm$. Supported by *ab-initio* calculations and atomistic spin dynamics simulations, we interpret this effect as the result of a reduced exchange coupling near the layer interfaces, which becomes significant as confinement increases. Two additional experiments were conducted on these samples to support this scenario: ultrafast linear X-ray transmission which is sensitive to photoinduced changes in the charge carrier population, and ultrafast electron diffraction (UED) which is sensitive to the photoinduced phonons.

## 2. Ultrafast X-ray Transmission: Probing Spins and Charge Carriers

A series of multilayer samples was synthesized in which the total Fe thickness is constant ($\sum d_{Fe} = 16$ nm). The samples differ in that they each have a different number of MgO spacer layers sectioning the total Fe thickness into thinner Fe layers of equal thickness $d_{Fe}$. For example, samples containing 1 or 7 spacer layers have $d_{Fe}$ = 8 nm or 2 nm, respectively (see Figure 1a). The MgO spacers are all 2 nm thick, and they are not excited by the pump laser pulses due to the high band gap of MgO (~7.8 eV). This spacer thickness does not allow charge or spin transport between adjacent Fe layers [12].

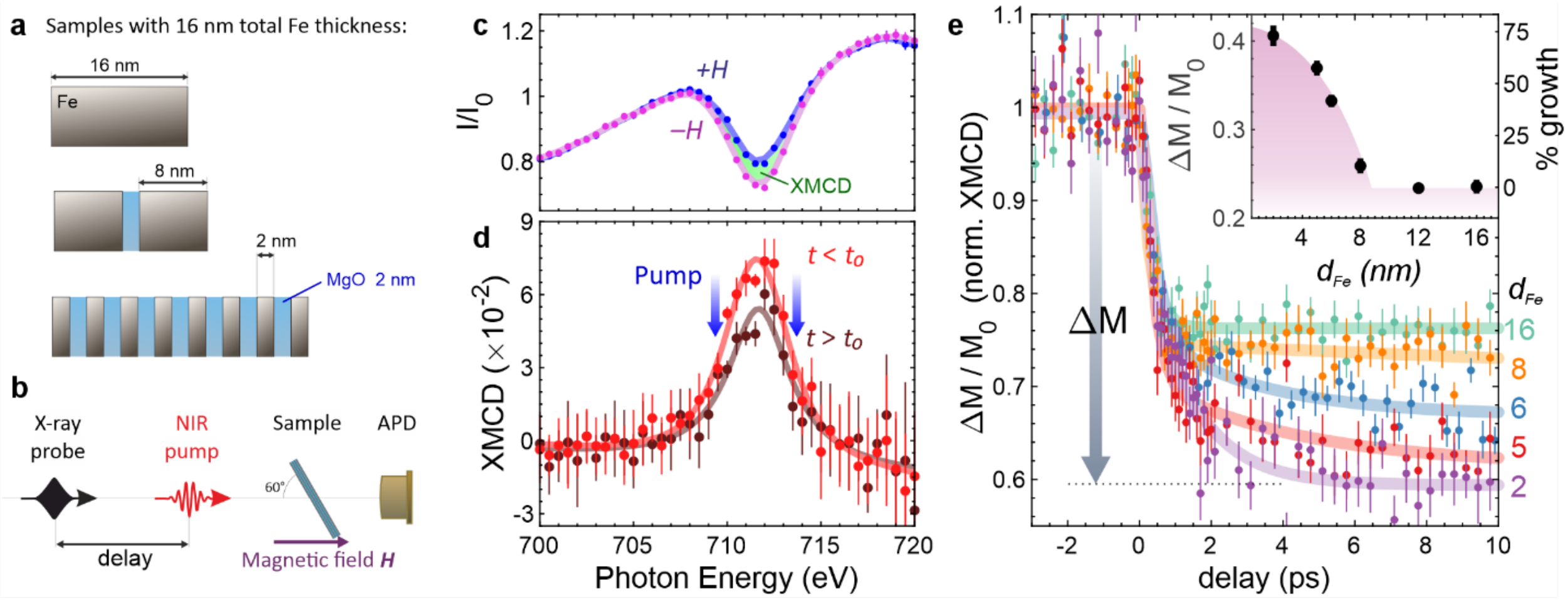

**Figure 1 – ultrafast transmission XMCD on Fe/MgO samples. a** Sketch of Fe/MgO multilayers with the same total Fe thickness (16 nm) but with a varying number of 2 nm MgO spacer layers. Each sample has a different Fe layer thickness $d_{Fe}$ (16, 8, and 2 nm are shown). **b** Sketch of the X-ray transmission experiment: the NIR pump and soft X-ray probe arrive collinearly at 60° to the sample surface. Magnetic fields are applied along the beams' propagation axis. **c** Transmission spectra collected from the $d_{Fe} = 5$ nm sample using circularly polarized X-rays near the Fe $L_3$ edge, under magnetic fields of ±0.4 T. The difference between the curves is XMCD. **d** The XMCD spectrum from **c**, shown before- and ~8.5 ps after the arrival of the pump pulse, exhibiting pump-induced suppression. **e** normalized transmission XMCD, representing normalized magnetization $M/M_0$ (see methods), as function of pump-probe delay from several Fe/MgO multilayer samples. The labels indicate the Fe layers' thickness $d_{Fe}$ in nm. $\Delta M$ is indicated for clarity. $\Delta M$ values from samples with total Fe thickness other than 16 nm were scaled to match the excitation density in samples with 16 nm total Fe thickness (see methods). Inset: demagnetization amplitude as functions of the Fe layer thickness (average of data between 8 and 10 ps). The right-hand vertical axis denotes % growth of $\Delta M$ from its bulk value. When error bars are absent, they are narrower than the marker.

We conducted X-ray transmission experiments at the Femto-Slicing facility [13,14] at BESSY II (see Figure 1b), using femtosecond circularly polarized X-ray pulses with energies at the Fe $L_3$ edge. Magnetic contrast through circular dichroism (XMCD) is achieved by switching the sign of an external magnetic field applied along the X-rays' propagation axis (see Figure 1c). We excited all samples with a femtosecond IR pulse ("pump"; see methods), arriving nearly parallel to the X-rays. This causes rapid demagnetization to occur in all measured samples (Figure 1d and 1e). Since the same volume of Fe is photoexcited and resonantly probed, any differences between the demagnetization processes in different samples are only expected due to the varying number of spacer layers or interfaces.

Indeed, substantial differences are observed. The data in Figure 1e exhibit two distinct processes. The first is the well-known sub-picosecond (ps) demagnetization [15–17] which is identical in all samples, exhibiting an exponential time constant of ~0.3 ps. The second demagnetization process occurs over several ps. The amplitude of this process exhibits a dependence on $d_{Fe}$, such that thinner layers exhibit notably larger demagnetization amplitudes. The inset of Figure 1e presents the total demagnetization amplitude as a function of $d_{Fe}$. Note that data from samples with other *total* Fe thicknesses ($\sum d_{Fe} \neq 16$ nm) are also shown in Figure 1e. The amplitudes in these data are scaled to match the excitation density of the $\sum d_{Fe} = 16$ nm samples with very few assumptions (see methods) and appear to agree well with the trend. In particular, we find that the increase in amplitude appears below ~10 nm, reaching a ~75% increase at $d_{Fe} = 2$ nm. This result, as presented in the inset of Figure 1e, is the primary experimental observation in this work and will be discussed further below.

To assist in interpreting this effect, we change the X-ray polarization to linear $\sigma$ (vertical) and keep the rest of the experiment unchanged. In this configuration the magnetic contrast is lost, and pump-induced changes to the transmission signal reflect a response dominated by the charge carrier population of the electronic structure near the Fermi level. [18,19] This approach is closely related to near-edge X-ray absorption spectroscopy (XAS)[19], and has been previously used to study such Fe/MgO heterostructures. [20] The XAS signals (Figure 2) exhibit a rapid drop followed by a recovery, with no discernable trend between samples of different $d_{Fe}$. These processes occur within ~1 ps. The time scale observed agrees with the duration of the first demagnetization step (Figure 1e).

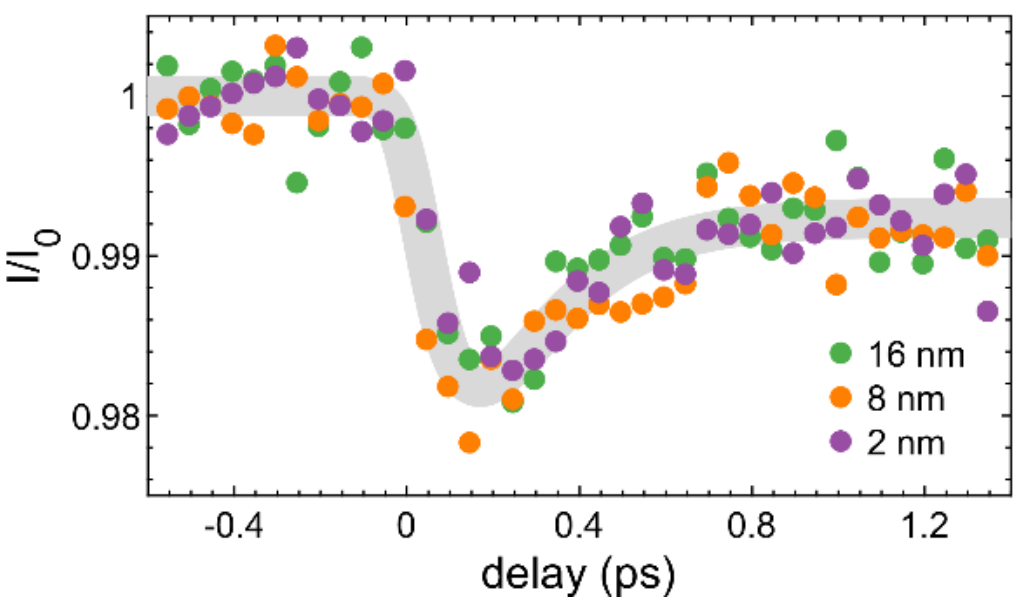

**Figure 2 – photoinduced reduction in transmission of linearly polarized X-rays at the Fe $L_3$ edge.** Data from three samples are shown, all with a total Fe thickness of 16 nm. The line is a guide to the eye.

## 3. Probing ultrafast phonon generation

For a broader picture of ultrafast dynamics in the confined Fe, we now turn to the phonon system. An ultrafast electron diffraction experiment (UED, see Figure 3a) was conducted at the SLAC MeV-UED facility. [21] UED provides direct access to the photoinduced changes in the lattice, [22] and has been previously employed to study lattice dynamics upon demagnetization. [20,23–27] As before, we study Fe/MgO samples in which we keep the total Fe thickness constant. Since MgO has a large band gap and is diamagnetic, the confined Fe layers only couple to it through interfacial vibrations. [12] To specifically address the effect of varying the amount of interfaces, we study two samples that contain the same amount of Fe and of MgO but differ in the number of Fe-MgO interfaces: one is a "bilayer" with 12 nm layers of Fe and MgO (i.e. 1 Fe-MgO interface) and the other is a "multilayer" with 6 repetitions of 2 nm Fe and 2 nm MgO (i.e. 11 Fe-MgO interfaces). As before, only the Fe is optically excited, and the MgO is not.

Debye-Scherrer diffraction ring patterns (seen in Figure 3a) were collected at every pump probe delay. At each delay, the pattern was azimuthally averaged to produce an intensity profile $I(q)$ as function of momentum transfer $q$ (Figure 3b).

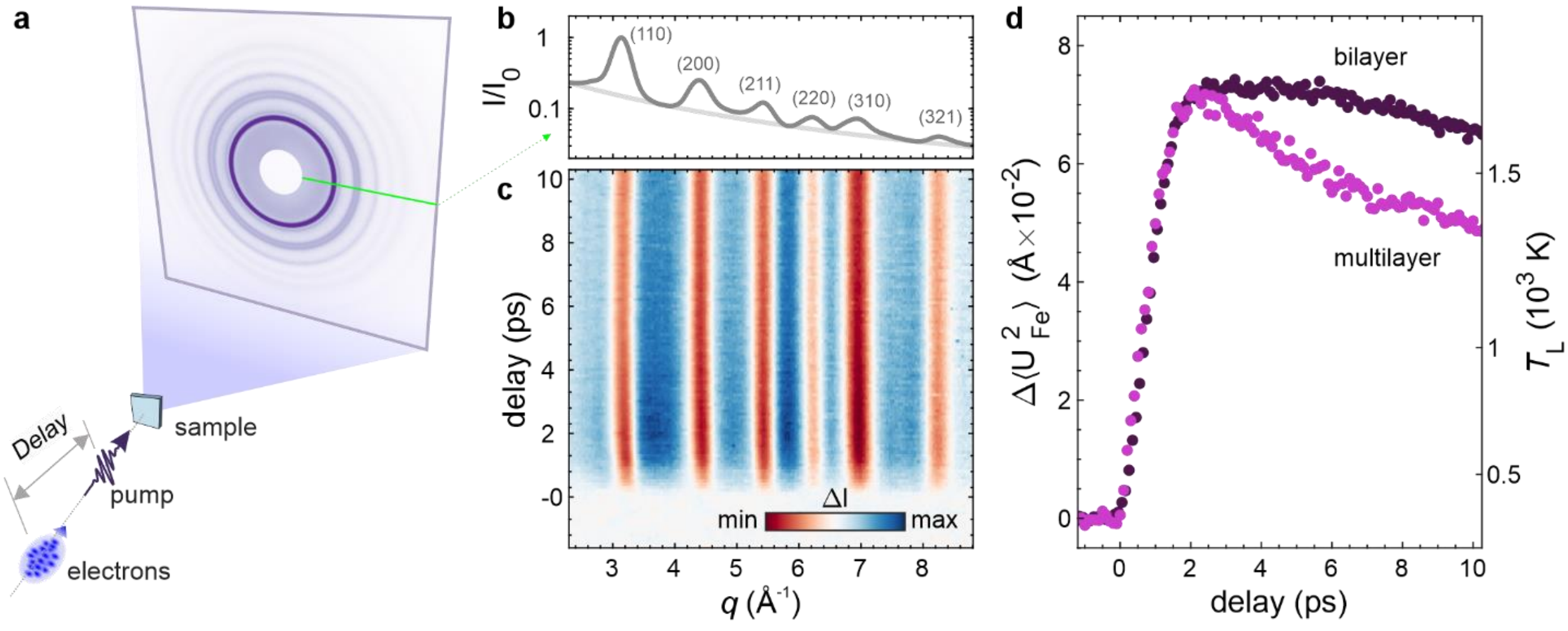

**Figure 3 – ultrafast electron diffraction**. **a** Sketch of the UED experiment. An example ring pattern from the experimental data is presented. **b** Azimuthal average of a ring pattern, normalized to the strongest peak. The monotonous curve illustrates the varying baseline. **c** Relative changes $\Delta I$ to the diffraction intensity pattern $I$ as function of momentum transfer $q$ and pump-probe delay. Data are taken from the multilayer sample. **d** Delay-dependent changes in Fe atoms' vibrational amplitude (mean squared displacement, MSD). The corresponding Fe lattice temperature is presented on the right hand axis. Two samples are measured under identical conditions, and both have the same material content: light symbols represent a $[Fe/MgO]_6$ multilayer of 2 nm individual layers, and dark circles represent an Fe/MgO bilayer with 12 nm layers. All data were taken with an incident fluence of 12 mJ cm$^{-2}$. Due to the high similarity between the Fe and MgO unit cells, most rings are a superposition of intensity from both materials, though only Fe is photoexcited. The data in **d** are extracted from rings that have no MgO contribution (see supplement).

Figure 3c presents all changes exhibited by $I(q)$ as a function of pump-probe delay and of $q$. We interpret all changes as due to a photoinduced rise in the phonon population: peaks weaken due to the Debye-Waller (DW) effect, and the diffuse scattering between them rises due to increased inelastic scattering from phonons. Peak weakening allows quantifying the increase in the atom's vibrations by means of their "mean squared displacement" (MSD, see methods).

Figure 3d presents the change in the Fe's MSD as a function of pump-probe delay from both samples. The data were collected from both samples under identical conditions, and both exhibited a nearly identical sharp rise in MSD within ~2 ps (in agreement with previous reports on elemental Fe [23]). This slightly exceeds the time frames of the XAS dynamics (Figure 2) and of the sub-ps demagnetization (Figure 1e). At delays beyond 2 ps, both MSDs begin to recover. However unlike their MSD rise, the two samples differ significantly in their MSDs' recovery, with the Fe lattice cooling down faster in the multilayer sample.

The identical MSD rises imply that the same amount of energy is transferred to the lattice in both samples, suggesting that the increased number of Fe-MgO interfaces does not markedly change the absorption of the pump beam. However, the accelerated recovery in the multilayer indicates that these interfaces play a significant role in transferring energy away from the excited Fe volume, most likely through interlayer vibrational coupling. [12]. We note that this recovery occurs concomitantly with the thickness dependent demagnetization process (Figure 1e).

The MSD data can be converted into units of degrees Kelvin using tabulated DW factors from literature, [28] as shown in the right-hand vertical axis of Figure 3d (see methods). This reflects the effective temperature of the Fe lattice $T_L$ under the assumption that increased phonon population is thermally distributed. Recent work on similarly simple metals such as Ni [25,27] and Sn [29] has shown that non-thermal phonon distributions can exists even several ps after the excitation. [12] We therefore stress that these numbers are reasonable figures of merit, but caution is needed before further interpretation.

## 4. Interpretation and *ab-initio* calculations

We begin by recalling that in the XAS experiment we observe no $d_{Fe}$-dependent trend in the time scales, and in the UED experiment we observe the same photoinduced growth in $\Delta\langle U_{Fe}^2\rangle$ (lattice vibrations) from both samples. Therefore, these experiments support the conclusion that pump absorption is not significantly affected by $d_{Fe}$. This indicates that the rise in demagnetization amplitude upon decreasing Fe thickness (Figure 1e) is magnetic in origin, and not an absorption effect (e.g. due to successive back reflections). Furthermore, the faster recovery of the multilayer in the UED experiment indicates that this magnetic effect is not phonon-driven, as that would suggest a reversed trend in Figure 1e. Atomistic spin dynamics simulations qualitatively reproduce our experimental results and support this conclusion as well (see supplement).

The primary effect of the increased Fe confinement is that spins near the interface have less neighbors to couple to, thereby locally weakening the spin order and magnetization, as sketched in Figure 4a for an Fe slab that is 8 atomic layers thick (2 nm). This effect weakens with distance from the interface, but as $d_{Fe}$ decreases the fraction of total Fe volume with weakened spin order may become significant. In this scenario, the same excitation density would result in a stronger effective demagnetization, explaining the effect in Figure 1e.

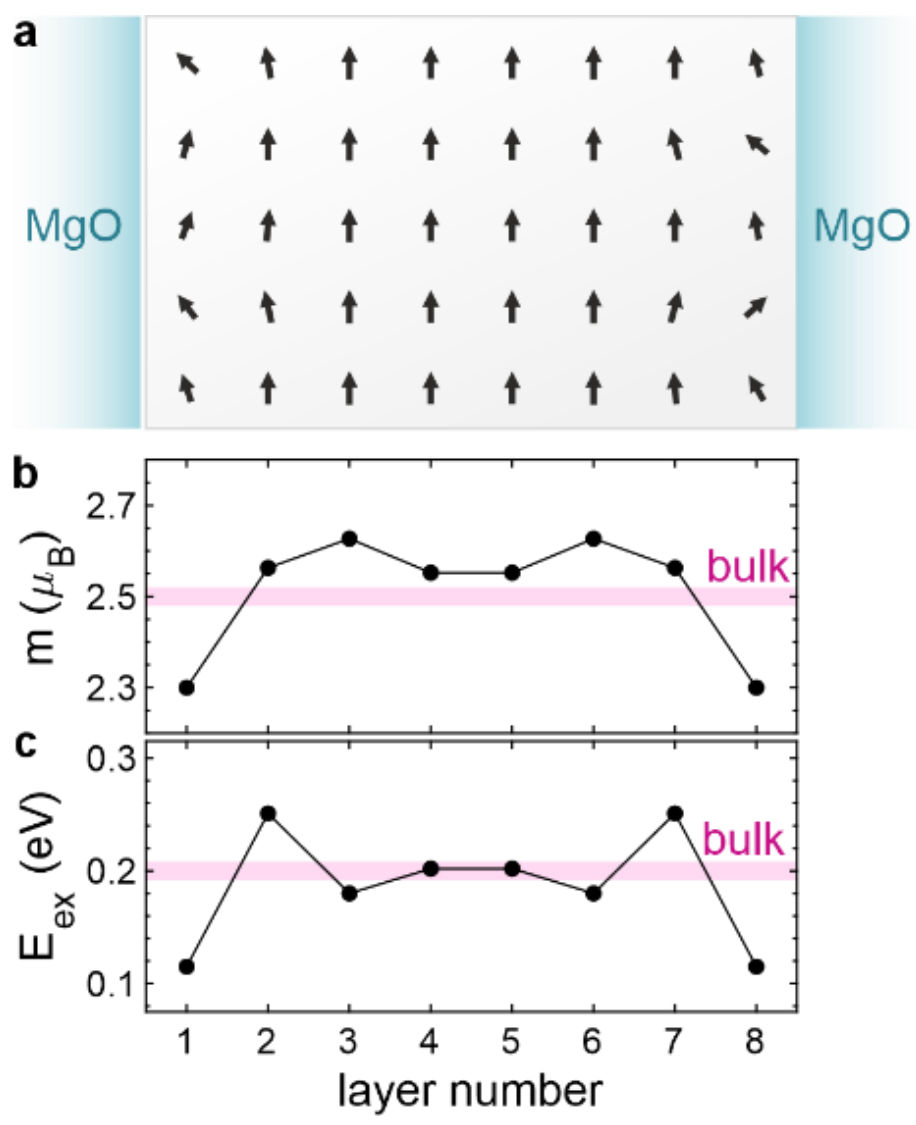

**Figure 4 - *ab-initio* calculation of a 2 nm Fe slab (8 atomic layers). a** Cartoon depiction of the Fe slab, highlighting a weakened spin order at its interfaces with MgO. **b** and **c** present the magnetic moment per Fe atom and the accumulated magnetic exchange energy (between an Fe atom and its environment, see methods) as functions of atomic layer number. Horizontal lines represent calculated bulk values for epitaxially grown Fe.

To estimate the spatial profile of magnetic interactions in a finite Fe layer, *ab-initio* calculations were conducted for an Fe slab that is 8 atomic layers thick (2 nm). Figs. 4b,c present the layer-specific magnetic moment per Fe atom and accumulated magnetic exchange energy, which is proportional to the mean-field Curie temperature in a corresponding bulk system. We find significant changes of over 50% in the latter, implying that the interfacial magnetization is significantly more susceptible to excitations despite only minor changes in the magnetic moments. Only the first atomic layer is significantly weakened by the interface, and by the third atomic layer the exchange energy is close to bulk values. This is comparable to the expected behavior of ultrathin ferromagnets in the mean field limit. [30] For such a 2 nm thick slab, this represents a weakening in 25% of the total volume. For $d_{Fe} = 10\ nm$, this reduces to only ~5%, suggesting that at higher $d_{Fe}$ values the effect is insignificant, as found in our XMCD experiment (Figure 1e). Importantly, such arguments are valid also for the present case of polycrystalline samples, as direct exchange is governed by local nearest-neighbor interactions.

Studies on Fe/MgO multilayers have shown that collective electronic effects can depend on similar microscopic effects, most notably the interlayer exchange field [31], which can depend on characteristics such as the number of repetitions N [32], and can favor antiferromagnetic alignment of layers. However our study investigated demagnetization occurring within a few ps and, starting from a FM arrangement. Excitations on this time scale are controlled by direct exchange (between neighboring Fe ions), and involve substantially higher energies compared to interlayer exchange (or dipolar coupling). This has recently been shown to be susceptible to MgO thickness and temperature [33,34], and may cause secondary processes on longer time scales.

**5. Conclusions**

We studied how nm-scale confinement affects the ultrafast demagnetization of Fe. A key challenge overcome in this study was to reliably maintain the same excitation and probing conditions while varying the level of confinement. This was done by fabricating a series of samples in which we kept the total amount of Fe constant and introduced a varying number of MgO spacers which serve to divide the Fe into thinner layers. Crucially, MgO is transparent to the excitation pulses and is not resonant with the transmission X-ray probes, thus ensuring comparable excitation and probing conditions. To study demagnetization, we used the same excitation on all samples and found that Fe layers with thicknesses below ~10 nm exhibit enlarged demagnetization amplitudes. In particular, we found that the amplitude increases as thickness is reduced, reaching a ~75% increase in 2 nm thick layers compared to bulk. Two additional experiments were conducted to test this effect. The transmission of linearly polarized X-rays (XAS, a probe of Fe charge carriers) exhibited no confinement-dependent dynamics. Also in ultrafast electron diffraction, a probe of phonons, we found no confinement-dependent effect in the heating of the Fe lattice. We found that lattice cooling occurs concomitantly to the amplified demagnetization effect and is accelerated when more Fe-MgO interfaces are present, indicating that the amplified demagnetization is not phonon-driven. Both experiments support the conclusion that the increase in demagnetization amplitude upon increased confinement is magnetic in origin. We therefore present a scenario in which the dominant effect of confinement is a weakening of spin order near the Fe-MgO interfaces. The ultrafast response of this "weakened" magnetic volume is negligible in bulk but becomes significant below ~10 nm. *Ab-initio* calculations were used to estimate the volume ratio of the interfacial layer, and found that for 10 nm Fe thickness this represents only 5% of the Fe, while for 2 nm the affected volume already reaches 25%. Therefore, we conclude that at such thicknesses the weakened interfacial Fe layers significantly influence the macroscopic response of the whole Fe layer.

Understanding how dimensional confinement affects the performance of ultrafast spin dynamics is essential for any realistic utilization in nm-scale devices, so results such as those presented here are of great need. The scenario we present suggests a route towards optimally harnessing FM confinement, such that the weakening of spin order does not compromise the FM order's stability in ambient conditions, but does allow using weaker excitation, which translates to more efficient energy consumption. Nevertheless, a deeper understanding of the effect is desirable.

## 6. Methods

*Sample Preparation:*

Polycrystalline heterostructure $[Fe/MgO]_n$ samples, where $n$ is the number of repetitions, were grown by molecular beam epitaxy under ultrahigh vacuum conditions on $Si_3N_4$ membranes, described in more detail elsewhere. [12,20] The MgO layers are always 2 nm thick. The Fe layer thickness $d_{Fe}$ is different for each sample, but in all samples the total Fe thickness across all layers is constant, fulfilling $n \cdot d_{Fe} = 16\,nm$. This defines the number of repetitions $n$ in each sample (e.g. the sample with $d_{Fe} = 8\,nm$ has $n = 2$ repetitions).

Three additional samples were used, for which $n \cdot d_{Fe} \neq 16\,nm$. These were: $[Fe/MgO]_3$ with $d_{Fe} = 5\,nm$ or with $d_{Fe} = 6\,nm$, and $[Fe/MgO]_2$ with $d_{Fe} = 12\,nm$. All samples used in the X-ray experiments had an additional 100 nm Cu layer below the $Si_3N_4$ membranes as a heat sink.

For UED experiments two samples were used. Both samples contain the same thickness of Fe and MgO (12 nm each), but differ in the distribution. These were a multilayer sample of [Fe (2 nm) / MgO (2 nm)]$_6$, and a bilayer sample of [Fe (12 nm) / MgO (12 nm)]$_1$.

The Fe/MgO interfacial structure was investigated using conversion electron Mössbauer spectroscopy. The interdiffusion across the Fe/MgO interface was shown to be below 0.2 nm, and is thus limited to a single atomic monolayer. For details see the supplementary material published with ref. [12].

*Ultrafast X-ray Transmission Experiments*

The ultrafast x-ray studies were conducted at the Femto-Slicing facility "FemtoSpeX" at the synchrotron light source BESSY II. X-ray pulses with photon energies near the Fe $L_3$ absorption edge were transmitted through the samples and recorded using an avalanche photodiode. The X-ray polarization was either circularly polarized (for XMCD contrast) or $\sigma$ linearly polarized (for probing charge carrier dynamics). Laser pulses of 1.55 eV photon energy were used to excite the sample. Both the X-rays and the laser pulses impinge on the sample surface at an angle of 60° to the surface. Transmission intensities were then collected as function of pump-probe delay $t$. During the XMCD scans, external magnetic fields of $\pm 0.4\,T$ were applied parallel to the X-ray beam to saturate the sample's magnetization $M$ (denoted at saturation as $M_0$), and intensities were collected for both field directions at every delay point. The data presented in Figure 1e are the difference between transmitted intensity under opposite fields $I_+ - I_-$, normalized by the value of $I_+ - I_-$ before the excitation. The assumption applied to these data is

$$I_+ - I_- \propto \exp(-\mu_+ d_{Fe}^{tot}) - \exp(-\mu_- d_{Fe}^{tot}) \approx -d_{Fe}^{tot}(\mu_+ - \mu_-) \propto \frac{M}{M_0}. \quad (1)$$

Here $d_{Fe}^{tot}$ is the total Fe thickness, and $\mu_\pm$ are the absorption lengths for opposite fields. The incident laser excitation fluence was set to 40 mJ/cm$^2$. Neglecting the effect of repeated back-reflections, this results in the same excitation density for all samples with $n \cdot d_{Fe} = 16\,nm$, because the MgO spacers are transparent to 1.55 eV. Using the absorption coefficient 4.2443×10⁵ cm⁻¹ and the lattice constant 2.866 Å, we calculate an effective excitation density of $\rho_0 = 1.1683$ photons per unit cell (see further details in the supplement). Figure 1e also presented XMCD results from additional samples with $n \cdot d_{Fe} \neq 16\ nm$. For the samples with $d_{Fe} = 5$, 6, and 12 nm the excitation was slightly different, resulting in excitation densities of $\rho = 1.1906$, 1.2745, and 1.6076 photons per unit cell, respectively. For a reliable comparison, the normalized XMCD data from these 3 samples were corrected before presentation in Figure 1e according to

$$I'(t) = \frac{\rho}{\rho_0}(I(t) - 1) + 1 \qquad (2)$$

Note that this correction affects amplitudes, not time scales. The $d_{Fe} = 12\ nm$ data can be found in the supplement.

The temporal resolution of the X-ray transmission experiment is estimated to be below 120 fs.

*Ultrafast Electron Diffraction*

The UED measurements were carried out at the SLAC MeV-UED facility in transmission geometry on two samples with the same total Fe and MgO thicknesses (see above). The samples were excited with 4.7 eV laser pulses of 50 femtosecond duration, with an overall incident fluence of 12 mJ/cm². They were subsequently probed by 200 femtosecond long electron pulses with 3.4 MeV kinetic energy. Both pulses arrive normal to the sample surface. The transmitted electrons were collected using a 2D detector, with which scattering intensity was recorded up to a momentum transfer of $q = 9$ Å$^{-1}$ (defined as $q = 2\pi \sin\theta / \lambda$, in which $\theta$ is the Bragg angle and $\lambda$ is the wavelength). A 2D scattering pattern was collected as function of pump-probe delay $\Delta t$. The scattered electrons produced a Debye-Scherrer ring pattern (as in Figure 3a) for every delay point $t$, which were then azimuthally averaged to produce scattered intensity profiles $I(q, t)$, as in Figure 3b. At momentum transfer values equal to those of Bragg reflections $Q_{(hkl)}$ ($h, k$, and $l$ are miller indices) the intensity can be converted into changes in atomic mean squared displacement $\Delta\langle u^2\rangle$ by following:

$$\frac{I(q,t)}{I(q,t<0)} = \exp\left(-\frac{1}{3}q^2\Delta\langle u^2\rangle(t)\right). \qquad \text{(M3)}$$

Due to the similarity between the unit cell dimensions of Fe and MgO, most rings from both materials overlap. Dynamics originating only from the Fe are extracted from the (211) and (321) rings, in which the contribution of MgO is negligible (see supplement). All experimental data were recorded at room temperature.

Lastly, the change in Fe MSD in Figure 3d was converted into units of temperature using tabulated DW factors from literature. [35] The tabulation uses a fourth order polynomial to express the DW factor $B = (8\pi^2/3)\langle u^2\rangle$ as function of temperature. We use a fourth order polynomial to express the reverse conversion following

$$T_L = 300\ K + \sum_{N=1}^{4} X_N \cdot (\Delta\langle u^2\rangle)^N\ . \qquad \text{(M4)}$$

The $X_N$ parameters used were $X_1 = 2.314 \times 10^4$ $°K/Å^2$, $X_2 = 1.64 \times 10^5$ $(°K/Å^2)^{-2}$, $X_3 = -5.13 \times 10^6$ $(°K/Å^2)^{-3}$, and $X_4 = 2.78 \times 10^7$ $(°K/Å^2)^{-4}$.

*Density Functional Theory calculations*

To address layer resolved magnetic interactions, we calculated Heisenberg exchange interaction constants $J_{ij}$ for a $Fe_8/MgO_8(001)$ heterostructure. These define the interactions between different pairs of atoms *i* and *j* of all different chemical types and positions as a function of the distance $r_{ij}$ between them in terms of a classical Heisenberg model Hamiltonian

$$\mathcal{H}_{\text{mag}} = -\sum_{i \neq j} J_{ij} \vec{S_i} \cdot \vec{S_j}\,, \qquad \text{(M5)}$$

where $\vec{S}_i$ and $\vec{S}_j$ describe the magnitude and the orientation of the magnetic spin moments at sites $i$ and $j$. We obtained the $J_{ij}$ optimized from first principles in the framework of density functional theory (DFT) using Korringa-Kohn-Rostoker (KKR) scheme following the Liechtenstein approach [36] as implemented in the Munich SPR-KKR package. [37] The exchange-correlation functional was treated within the GGA PBE scheme in combination with the atomic sphere approximation (ASA). We used an angular momentum expansion reaching up to *f*-states ($l_{max} = 4$). Due to the presence of the insulator, Lloyd scaling is used to improve convergence. We assumed electronic self-consistency to be reached when the error in the potential functions dropped below $10^{-5}$. Brillouin zone integration was carried out with the help of the special point method using a regular $k$-point grid of 272 points in the irreducible Brillouin zone, which corresponds to a 31 × 31 × 3 mesh in the full Brillouin zone. Optimized lattice parameters and atomic positions of the cell, which consists of 4 inequivalent Fe, Mg and O sites, respectively, were taken from previous DFT calculations. [12]

To quantify the site-specific magnetic interactions, we summed up the effective magnetic exchange energy $J_{ij}\vec{S_i} \cdot \vec{S_j}$ for each site *i* over all neighbors *j* up to 4 lattice constants apart. In a bulk system with a single magnetic lattice site a positive sum is directly proportional to the Curie temperature $T_C$ in mean field approximation. For bcc bulk Fe calculated with a similar setup at the experimental lattice constant $a = 2.83$ Å, we obtained a spin moment of 2.29 $\mu_B$ and an accumulated exchange constant of 143 meV which corresponds to $T_C = 1106\,K$. In heterostructures, epitaxial match causes significant lattice strain, which affects magnetism as well. We obtain a substantially higher moment of 2.53 $\mu_B$. Here the accumulated exchange amounts to 172 meV, corresponding to a mean field value of $T_C = 1329\,K$.

## Acknowledgements

Funding by the Deutsche Forschungsgemeinschaft (DFG, German Research Foundation) by Project-ID 278162697 - SFB 1242, Project-ID 328545488 - SFB-TRR 227 (projects A10, A09, and A03), Project-ID 425217212 - SFB 1432, and the Emmy Noether program (Grant No. RE 3977/1) is gratefully acknowledged. The Helmholtz Zentrum Berlin is gratefully acknowledged for the allocation of synchrotron radiation beamtime and for financial support for travel to BESSY II. We acknowledge the beamtime support of Florin Radu and Torsten Kachel. The UED work was performed at SLAC MeV-UED, which is supported in part by the DOE BES SUF Division Accelerator & Detector R&D program, the LCLS Facility, and SLAC under Contracts No. DE-AC02-05-CH11231 and No. DE-AC02-76SF00515. Calculations were carried out on the MagnitUDE supercomputer system (DFG Grants No. INST 20876/209-1 FUGG and No. INST 20876/243-1 FUGG) of the Center for Computational Sciences and Simulation at the University of Duisburg-Essen. M.M. is supported by the U.S. Department of Energy Fusion Energy Sciences under FWP 100182. We thank R. Mitzner, K. Holldack, and J. Yang for experimental support. We further thank U. von Hörsten for his expert technical assistance with sample preparation, and Florian Denizer for early experimental tests. We acknowledge the Scientific Compute Cluster in Universität Konstanz (SCCKN) for providing the computational resources and support that contributed to the ASD results.